\documentstyle[12pt]{article}
\def\beq{\begin{equation}}
\def\eeq{\end{equation}}
\def\beqa{\begin{eqnarray}}
\def\eeqa{\end{eqnarray}}
\def\beq{\begin{equation}}
\def\eeq{\end{equation}}
\def\beqa{\begin{eqnarray}}
\def\eeqa{\end{eqnarray}}
\newcommand{\be}{\begin{equation}}
\newcommand{\eq}{\end{equation}}
\begin{document}
\begin{flushright}
\hfill QMW-TH-98/5\\
\hfill hep-th/9801161\\
\end{flushright}
\vspace{1cm}
\begin{center}
\baselineskip=16pt
{\Large\bf Metrics Admitting Killing Spinors In Five Dimensions} 
\vskip 2cm
{\bf A. H. Chamseddine}\\
\vskip 0.5cm
{\em Institute of Theoretical Physics,\\
ETH-H\"onggenberg, \\
CH-8093 Z\"urich, Switzerland}\\
E-mail: chams@itp.phys.ethz.ch
\vskip 0.8cm
{\bf W. A. Sabra}\\
\vskip 0.5cm
{\em Physics Department,\\
Queen Mary and Westfield College,\\
Mile End Road, E1 4NS}\\
E-mail: W.Sabra@qmw.ac.uk
\vskip 0.2cm
\end{center}
\vskip 1 cm
\begin{abstract}
BPS black hole configurations which break half of supersymmetry in  
the theory of $N=2$ $d=5$
supergravity coupled to an arbitrary number of abelian vector multiplets 
are discussed. A general class of solutions comprising all known BPS
rotating black hole solutions is obtained.
 
\end{abstract}
\bigskip
\bigskip
\newpage
\section{Introduction} 
Recently there has been lots of interest in the
study of BPS black hole solutions 
of the low-energy effective action of compactified string 
and $M$-theory. These activities have been initiated mainly 
due to the realisation that the recent 
understanding of the non-perturbative structure of string theory
provides the microscopic degrees of freedom, $D$-branes, 
which give rise to the 
Bekenstein-Hawking entropy. In particular, BPS saturated solutions of 
toroidally compactified string theory  have been constructed and their 
entropy was microscopically calculated using ``D-brane'' technology.
Later, the microscopic analysis was applied to near-extreme black hole 
solutions for both the static and the rotating case.
However, in these cases, the arguments become more heuristic 
and less rigorous (
for a review, see for example, \cite{peet}).

BPS saturated solutions in toroidal compactifications correspond 
to vacua with $N=4$ and $N=8$ supersymmetry. These are severely 
constrained by the large supersymmetry and corrections to these 
solutions and their entropies can only arise 
from higher loop corrections, as the lowest order corrections are
known to vanish.
In contrast, BPS saturated solutions of $N=2$ string vacua can 
receive corrections  even at the one loop level.
In the context of string theories, $N=2$ supergravity models in four and
five dimensions  with 
vector and hyper-multiplets 
arise, respectively, from type II string and M-theory 
 compactified on a Calabi-Yau
threefold. The presence of perturbative and nonperturbative corrections for 
these models makes $N=2$ black holes more intricate and also more 
difficult to analyse. However, the analysis of these black holes can 
be considerably simplified by the 
rich geometric structure of the underlying four and five dimensional 
low-energy effective field  theory provided, respectively, 
by special and 
very special geometry \cite{special, very}. For example, the metric in
four dimensions can be expressed in terms of
symplectic invariant quantities in which the symplectic sections 
satisfy algebraic constraints involving a set of constrained harmonic 
functions \cite{sabra1}. 
Like for Einstein-Maxwell theory \cite{HH}, various types of solutions, such as
rotating and TAUB-NUT spaces, depend very much on the 
choice of the harmonic functions as well as the prepotential defining 
special geometry.

Using the explicit static black hole solutions, one can calculate the entropy
and the value of the scalar fields near the horizon. 
An important feature of these
black holes is that, for those with non-singular horizons, the entropy
can be expressed in terms of the  extremum of the 
central charge and that the scalar fields take fixed values 
at the horizon independent 
of their initial values at spatial infinity \cite{feka}. 

In five dimensions, static metrics admitting supersymmetry or Killing spinors 
were constructed for the case
of pure $N=2$ supergravity in \cite{gkltt}. The metric in this case 
is of the Tanghelini form \cite{frt,perry}.
In the context of $N=2$ supergravity in five dimensions with abelian
vector multiplets, 
extreme black holes with constant scalars, the so-called double-extreme BPS 
black holes, were considered in
\cite{chams}. It was also shown that 
their entropy can be expressed in terms of the extremised central charge. 
Moreover, the Strominger-Vafa black hole \cite{SV} was reproduced as a 
double-extreme black hole of an $N=2$ supergravity model with one 
vector multiplet.

Moreover, an extremal rotating  
back hole solution was constructed in \cite{ppp}. This solution was later
embedded into $N=2$ supergravity theory in five dimensions interacting 
with one vector multiplet whose scalar field is set to a constant 
\cite{raja}. This solution was then shown to be supersymmetric by
solving for the Killing spinor equations.        
Rotating black hole solutions for five-dimensional $N=4$ superstring 
vacua were also constructed in \cite{mirjam},\cite{Tsey}.

It is our purpose in this work to study general 
BPS black hole solutions which break half of supersymmetry of  
$d=5,$ $N=2$ supergravity theory with an arbitrary number of vector multiplets.
Static non-rotating solutions have been discussed recently in
\cite{sabra2}. This will be generalised here to allow for rotating
solutions. The class of solutions obtained will include all known
rotating BPS black hole solutions.
This work is organised as follows.
In the next section, the structure of $d=5$, $N=2$ supergravity is
briefly reviewed, and we collect some  formulae and expressions 
which will be relevant for our later discussion.
In section three,  we will present  static black hole solutions and 
verify that they admit unbroken supersymmetry by solving for the
supersymmetry transformation rules for the gravitino and the gauginos in a
bosonic background. 
\section {$d=5$ $N=2$ Supergravity and Very Special Geometry}
The theory of $N=2$ supergravity coupled to an arbitrary 
number $n$ of Maxwell's supermultiplets was first considered in \cite{GST}.
In this work, it was established that the real scalar fields of 
the vector supermultiplets
parametrise a riemannian space. The classification 
of homogeneous symmetric spaces is related to that of Jordan algebras of
degree three. These spaces can be expressed in the form 
\begin{equation}
{\cal M}={\hbox{Str}_0(J)\over \hbox{Aut}(J)},
\end{equation}
where $\hbox{Str}_0(J)$ is the reduced structure group of a 
formally real unital Jordan Algebra, $\hbox{Aut}(J)$ is its automorphism 
group. The scalar manifold can be regarded as a hyperspace,
with vanishing second fundamental form of an $(n+1)$-dimensional riemannian
space $\cal G$ whose coordinates $X$ are in correspondence with the vector
multiplets including that of the graviphoton. The equation of 
the hypersurface is ${\cal V}=1$, where 
$\cal V$ , the prepotential, is a homogeneous cubic polynomial in the 
coordinates of $\cal G,$
\begin{equation}
{\cal V} = {1\over 6} C_{IJK} X^I X^J X^K.
\label{pin}
\end{equation}
More recent treatment of the bosonic part of  $N=2$ supergravity theory 
was given in \cite{very} in terms of ``very special geometry''. 
The construction of 
$N=2$ supergravity arising from the compactification of  $11$ dimensional 
supergravity on a Calabi-Yau 3-folds was discussed  more recently in
 \cite{Cadavid}. 

The bosonic part of the effective supersymmetric 
$N=2$ Lagrangian which describes the coupling of vector multiplets to
 supergravity is entirely determined in terms 
of the homogeneous cubic prepotential (\ref{pin}) defining very special 
geometry \cite{very} and which in the case of Calabi-Yau compactification
corresponds to the 
intersection form. This Lagrangian is   
\begin{equation}
e^{-1} {\cal L} = -{1\over 2} R - {1\over 4} G_{IJ} F_{\mu\nu} {}^I
F^{\mu\nu J}-{1\over 2} g_{ij} \partial_{\mu} \phi^i \partial^\mu \phi^j
+{e^{-1}\over 48} \epsilon^{\mu\nu\rho\sigma\lambda} C_{IJK} 
F_{\mu\nu}^IF_{\rho\sigma}^JA_\lambda^K
\label{action}
\end{equation}
where $R$ is the scalar curvature,$F_{\mu\nu} {}^I =
2\partial_{[\mu}A_{\nu]}^I$ 
is the Maxwell field-strength
tensor and $e=\sqrt{-g}$ is the determinant of the F\"unfbein 
$e_m^{\ a}$. 
\footnote {In this paper, we shall be using the signature $(-++++)$ 
and for the indices we take: $m,n,\cdots$ to denote curved indices  whereas 
the indices $a, b,\cdots$ are flat. 
Antisymmetrized indices are defined by: $[ab]
= {1 \over2} (ab -ba)$.}

The fields $X^I= X^I(\phi)$ are the
special coordinates satisfying
\begin{equation}
X^I X_I=1 , \qquad {1\over 6}C_{IJK} X^I X^J X^K =1,
\label{cn}
\end{equation}
where, $X_I$, the dual coordinate is defined by,
\begin{equation}
X_I={1\over6}C_{IJK}X^JX^K.
\label{d}
\end{equation}
The gauge coupling metric $G_{IJ}$ which depends on the moduli, and
the metric $g_{ij}$ 
are given in terms of 
the prepotential (\ref{pin}) by
\begin{eqnarray}
G_{IJ} &=& -{1\over 2}{\partial\over \partial X^I}
{\partial\over\partial X^J}(\ln {\cal V})|_{{\cal V} =1},\nonumber\\
g_{ij}& =& G_{IJ} \partial_{i}X^I\partial_{j}X^J|_{{\cal V} =1},\qquad  
(\partial_i \equiv {\partial \over \partial\phi^i}).
\label{metric}
\end{eqnarray}

Now we list some useful relations which follow from very special geometry.
Using the definition ${\cal V}=1$,
one can easily deduce that
\begin{equation}
\partial_iX_{I} = {1\over3}C_{IJK} \partial_iX^J X^K, \qquad
X^I\partial_iX_I=X_I\partial_iX^I=0.
\label{si}
\end{equation}
Moreover, using the definition of (\ref{metric}),  
the gauge coupling metric can be expressed 
in terms of the special coordinates by 
\begin{equation}
G_{IJ}=-{1\over2} C_{IJK}X^K+{9\over2}X_I X_J.
\label{gcs}
\end{equation}
Also one can easily verify the following relations
\begin{eqnarray}
X_I&=&{2\over3}G_{IJ} X^J,\nonumber\\
\partial_i X_I&=&-{2\over3}G_{IJ}\partial_iX^J.
\label{scg}
\end{eqnarray}
 
The supersymmetry transformation laws for the Fermi fields in a 
bosonic background are given by \cite{GST,chams}
\begin{eqnarray}
\delta\psi_\mu &=& {\cal D}_\mu\epsilon + { i\over 8}
X_I
\Bigl(\Gamma_\mu{}^{\nu\rho} - 4 \delta_\mu{}^ \nu \Gamma^\rho\Bigr)
F_{\nu\rho}{}^I \epsilon,\nonumber\\
\delta \lambda _i &=&{3\over 8}\partial_iX_I\Gamma^{\mu\nu}\epsilon
F_{\mu\nu}^I
- {i\over 2} g_{ij} \Gamma^\mu \partial_\mu\phi^j \epsilon,
\label{stl}
\end{eqnarray}
where 
$\epsilon$ is the supersymmetry parameter and ${\cal D}_\mu$ the covariant derivative
\begin{equation}
{\cal D}_\mu=\partial_\mu+{1\over4}\omega_{\mu ab}
\Gamma^{ab}.
\end{equation}
Here, $\omega_{\mu ab}$ is the spin connection, 
$\Gamma^{{\nu}}$ are Dirac matrices and 
\begin{equation}
\Gamma^{{a}_1{a}_2\cdots{a_n}}=
{1\over n!}\Gamma^{[{a_1}}\Gamma^{{a_2}}
\cdots \Gamma^{{a_n}]}.
\end{equation}
\section{BPS Rotating Black Hole Solutions}
We are interested in finding BPS black hole solutions which break 
half of the supersymmetry 
of the underlying $N=2$ $d=5$ supergravity in five dimensions. 
Our analysis is for a generic
model with arbitrary number of vector multiplets and general values 
for $C_{IJK}$. 
Motivated by the form of the non-rotating metric which admits supersymmetry 
in the case of 
pure supergravity with no vector multiplets \cite{gkltt} as well as the static 
solutions found in \cite{sabra2}, we assume that the metric
can be brought to the form
\begin{equation}
ds^2 =-e^{-4 U}(dt+w_mdx^m)^2 +e^{2 U} (d\vec{x})^2
\label{magic}
\end{equation}
where $U=U(x)$ and $w_m=w_m(x)$.
Note that for $w_m=0$ (no rotation), and for pure supergravity,
 the solution which admits 
Killing spinors is known to be given by \cite{gkltt} 
\begin{equation}
ds^2=-H^{-2}dt^2+H(d\vec x)^2,
\end{equation}
where $H$ is a harmonic function.

The F\"unfbeins for the metric in (\ref{magic}) are
\begin{eqnarray}
e_{{t}}^{\ 0} &=& e^{-2U} \quad , \quad 
e_{{m}}^{\ 0} = e^{-2U} w_m\nonumber \\
e_{{t}}^{\ a} &=& 0\quad ,\quad e_{{m}}^{\ a}=e^{U}\delta_m^a\nonumber\\ 
e^{{t}}_{\ 0} &=& e^{2U}\quad , \quad 
e^{{t}}_{\ a} = -e^{-U}w_m\delta_a^m\nonumber\\
e^{{m}}_{\ 0} &=& 0\quad , \quad 
e^{{m}}_{\ a} = e^{-U}\delta_a^m
\label{funf}
\end{eqnarray}
For the spin connections one obtains
\begin{eqnarray}
\omega_{t}^{a0} &=&2\partial_mU(e^{-3U})\delta_a^m,\nonumber\\
\omega_{n}^{a0} &=&2\partial_mUe^{-3U}w_n\delta_m^{\ a}
+{1\over2}e^{-3U}\delta^{am}(\partial_n w_m-\partial_m w_n)\nonumber\\
\omega_{t}^{ab} &=&-{1\over2} e^{-6U}
\delta_a^m\delta_b^n(\partial_nw_m-\partial_mw_n),\nonumber\\
\omega_{p}^{ab} &=&-{1\over2} e^{-6U}w_p\delta^{am}\delta^{bn}
(\partial_nw_m-\partial_mw_n)-
\partial_nU(\delta^{an}\delta^{b}_p-\delta^{nb}\delta^{a}_p)\nonumber\\
\end{eqnarray}

 We now turn to our Ansatz and try to 
 determine the constraints imposed by unbroken supersymmetry on
the functions $U$ and $w$. This is achieved by solving the equations
obtained by demanding the vanishing of the gravitino and gauginos supersymmetry
transformation laws in a bosonic background. 
We first start with the equation corresponding to the 
vanishing of the time
component of the gravitino supersymmetry transformation. For our
metric this  is given by
\begin{eqnarray}
\delta\psi_t&=&\partial_t\epsilon
-{1\over8}e^{-6U}\delta^{am}\delta^{bn}
\Gamma_{ab}\Big((\partial_n w_m-\partial_mw_n)
+e^{2U}X_IF^I_{mn}\Big)\epsilon\nonumber\\
&+&i
\partial_mU\delta^{am}e^{-3U}\Gamma_a\epsilon
+{1\over4}e^{-4U} w_m\delta^{am}\delta^{bn}X_IF^I_{tn}\Gamma_{ab}
\epsilon\nonumber\\
&-&{i\over2}e^{-2U}
X_IF_{tm}^I\Gamma^a\epsilon\delta_a^m.\nonumber
\end{eqnarray}
Using the relation 
\begin{equation}
\Gamma^{ab}=-{i\over2}\epsilon^{abcd0}\Gamma_{cd}\Gamma_0,
\end{equation}
and demanding that $\Gamma^0\epsilon=-i\epsilon$, 
we obtain the following relations for the supersymmetry transformation
parameter and the graviphoton field strength
\begin{eqnarray}
\partial_t\epsilon&=& 0,\nonumber\\
(X_IF^I_{mn})^-&=&(\partial_mQ_n-\partial_nQ_m)^-,\nonumber\\
(X_IF^I_{tm})&=&-\partial_me^{-2U}.
\label{firstansatz}
\end{eqnarray}
where $Q_n\equiv e^{-2U} w_n$ and $F^-_{mn}=F_{mn}- ^{*}F_{mn}.$
Notice that the chirality constraint on the spinor $\epsilon $ reduces
the $N=2$ supersymmetry to $N=1$.
Next, we turn to the space-component of the gravitino supersymmetry
 transformation. 
Using our Ansatz, we obtain the following equation
\begin{eqnarray}
\delta\psi_m&=&\partial_m\epsilon
+{1\over4}(\omega_m^{ab}\Gamma^{ab}+2\omega_m^{a0}
\Gamma_a\Gamma_0)\epsilon\nonumber\\
&+&{i\over8}X_I(\Gamma_m^{\ np}F_{np}^I+2\Gamma_m^{\ nt}F_{nt}^I-
4\Gamma^nF_{mn}^I-4\Gamma^tF_{mt}^I)\epsilon=0
\end{eqnarray}
The $\Gamma_a$ dependent terms are independent of the other terms and
have to vanish, this implies 
the condition
\begin{eqnarray}
& & 2(\partial_mUw_n-\partial_nUw_m)
-{1\over2}(\partial_mw_n-\partial_nw_m)+\nonumber\\
& &e^{2U}(X_IF_{mq}^I)-{1\over2}
e^{2U}(X_IF_{mq}^I)^*-(\partial_mUw_n-\partial_nUw_m)^*=0
\end{eqnarray}
The above equations together with (\ref{firstansatz}) then gives  
\begin{eqnarray}
(\partial_mw_n-\partial_nw_m)^-&=&0\nonumber\\
X_IF^I_{mn}&=&(\partial_mQ_n-\partial_nQ_m).
\label{juliet}
\end{eqnarray}
Using the above relations, it can be easily shown that the 
$\Gamma_{ab}$ coefficient is identically vanishing.
Therefore, the vanishing of the space-component of the 
gravitino transformation, with the supersymmetry breaking condition, amounts
to the following simple differential equation,
\begin{equation}
(\partial_m+\partial_mU)\epsilon=0
\end{equation}
which admits the solution
\begin{equation}
\epsilon=e^{-U}\epsilon_0
\end{equation}
where $\epsilon_0$ is a constant spinor satisfying 
$\Gamma^0\epsilon_0=-i\epsilon_0.$
Finally we consider the supersymmetry transformation law of the gauginos given 
in (\ref{stl}). Using
(\ref{metric}), this can be rewritten in the form 
\begin{equation}
\delta\lambda_i=-{1\over4}\Big(G_{IJ}\partial_iX^I\Gamma^{\mu\nu}F^J_{\mu\nu}-
{3i}\Gamma^\mu\partial_\mu X_I\partial_iX^I\Big)\epsilon
\end{equation}
The vanishing of the gaugino transformation thus gives
\begin{equation}
G_{IJ}\partial_iX^I(\Gamma^{mn}F_{mn}^I+
2\Gamma^{mt}F_{mt}^I)\epsilon
-{3i}\partial_\mu X_I\partial_iX^I\Gamma^m\epsilon=0
\end{equation}

This leads to the two equations corresponding to the vanishing of the 
coefficients of the $\Gamma^m $ and $\Gamma^{mn} $ terms  
\begin{eqnarray}
{3\over2}e^{-U}\partial_mX_I\partial_iX^I
-G_{IJ}e^U\partial_iX^IF_{tm}^J&=&0, \nonumber\\
\Big(G_{IJ}\partial_iX^IF_{mn}^I+
{3\over2}e^{-2U}(\partial_mX_I w_n-\partial_nX_I w_m)\Big)^-=&0&.
\label{geq}
\end{eqnarray}
The first equation can be solved by 
\begin{equation}
F_{tm}^I=-\partial_m(e^{-2U}X^I).
\label{grace}
\end{equation}
This can be verified by noticing that  
\begin{eqnarray}
\partial_iX^IG_{IJ}F_{tm}^J&=& 
-G_{IJ}\partial_iX^I\partial_me^{-2U}X^J-G_{IJ}\partial_iX^Ie^{-2U}
\partial_mX^J
\nonumber\\
 &=&{3\over2}e^{-2U}\partial_mX_I\partial_iX^I
\label{pk}
\end{eqnarray}
where we have made use of the relations in (\ref{si}) and (\ref{scg}).
Clearly the Ansatz (\ref{grace}) is consistent with 
(\ref{firstansatz}),
\begin{equation}
X_IF^I_{tm}=-X_I\partial_m(e^{-2U}X^I)=-\partial_m(e^{-2U})
\end{equation}
where we have made use of the relation $X^IX_I=1.$
Also one can easily verify that
\begin{equation}
G_{IJ}F_{tm}^J={3\over2}e^{-4U}\partial_m(e^{2U}X_I)
\end{equation}

The second equation in (\ref{geq}) gives
\begin{equation}
G_{IJ}\partial_iX^I(F_{mn}^J)^{-}=-{3\over2}
e^{-2U}\partial_iX^I(\partial_mX_IQ_n-\partial_nX_IQ_m)^{-}.
\label{gg}
\end{equation}
This together with (\ref{juliet}) can be solved
by  
\begin{equation}
F^I_{mn} =\partial_m(X^IQ_n)-\partial_n(X^IQ_m).
\label{mona}
\end{equation}

In order to fix the various quantities in terms of space-time functions,
we solve for the equations of motion for the gauge fields. From the 
Lagrangian (\ref{action}), 
one can derive the following equation of motion for the gauge fields,
\footnote{notice that  
the Bianchi identities are trivially satisfied}
\begin{equation}
\partial_\nu(eG_{IJ}g^{\mu\rho}g^{\nu\sigma}
F_{\rho\sigma}^J)={1\over 16}C_{IJK}\epsilon^{\mu\nu\rho\sigma k}
F_{\nu\rho}^JF_{\sigma k}^K.
\label{gaugeeqn}
\end{equation}
After some lengthy calculation, one finds that for our Ansatz the 
equation of motion 
(\ref{gaugeeqn}) gives
\begin{equation}
e^{2U}X_I={1\over 3}H_I.
\label{jennie}
\end{equation}
where $H_I$ is a harmonic function.
Using this algebraic equation together with (\ref{cn}) and (\ref{d}),
one can determine $X^I$ and the metric in terms
of a set of harmonic functions. The gauge fields are then determined
by using their relations to the special coordinates given by
(\ref{grace}) and (\ref{mona}).
Before we discuss particular solutions, we will demonstrate that the
BPS
 solution discussed can be expressed
in terms of the geometry of the internal space, i. e., can be related
to the cubic polynomial $\cal V$.
If we define the rescaled coordinates 
\begin{equation}
Y_I=e^{2U}X_I, \qquad Y^I=e^{U}X^I,
\end{equation}
then the underlying very special geometry implies that
\begin{equation}
Y_IY^I=e^{3U}X_IX^I=e^{3U}={\cal V}(Y)={1\over3}C_{IJK}Y^IY^JY^K
\end{equation}
and thus the metric takes the form 
\begin{equation}
ds^2=-{\cal V}^{-4/3}(Y)(dt+\omega_mdx^m)^2+{\cal V}^{2/3}(Y)(d\vec{x})^2
\end{equation}
where  
\begin{equation}
{1\over2} C_{IJK}Y^JY^K=H_I.
\end{equation}

Using the above general Ansatz one can construct models 
with rotational symmetry. 
To proceed perhaps it is more convenient to work in  
spherical coordinates.
If we choose the four spatial coordinates as
\begin{equation}
x^1+ix^2=r\sin\theta e^{i\phi},\qquad x^3+ix^4=r\cos\theta e^{i\psi}
\end{equation}
and specialise to solutions with rotational
symmetry in  two orthogonal planes, i. e.,
\begin{equation}
w_\phi=w_\phi(r, \theta),\qquad w_\psi=w_\psi(r, \theta),
\qquad w_r=\omega_\theta=0.
\end{equation} 
then the  self-duality condition of the field strength of $w$
implies
\begin{eqnarray}
\partial_r w_\phi+{\tan\theta\over r}\partial_\theta w_\psi&=&0,
\nonumber\\
\partial_\theta w_\phi-r\tan\theta\partial_r w_\psi &=&0.
\end{eqnarray}
One finds for a (decaying) solution,
\begin{equation}
w_\phi=-{\alpha\over r^2}\sin^2\theta,\qquad w_\psi=
{\alpha\over r^2}\cos^2\theta,
\end{equation}
which in Cartesian coordinates give
\begin{equation}
w_1={\alpha x^2\over r^4},\qquad 
w_2=-{\alpha x^1\over r^4},\qquad
w_3=-{\alpha x^4\over r^4},\qquad
w_1={\alpha x^3\over r^4}.
\end{equation}
Thus the angular momentum is given by 
\cite{perry}
\begin{equation}
J^{(12)}=J^\phi=-J^{(34)}=-J^\psi={\alpha\pi\over 4G_N}.
\end{equation}
Therefore, the general form of solution with rotational symmetry in
two orthogonal planes has  the following form for the metric
\begin{equation}
ds^2=-e^{-4U}\Big(dt-{\alpha\sin^2\theta\over r^2} d\phi+
{\alpha\cos^2\theta\over r^2}\Big)^2+e^{2U}\Big(dr^2+r^2d\Omega^2)
\end{equation}
where 
\begin{equation}
d\Omega^2=(d\theta^2+\sin^2\theta d\phi^2+\cos^2\theta d\psi^2).
\end{equation}
Let us examine the behaviour of our solution near the horizon,
($r\rightarrow 0$).
There ${\cal V}(Y)$ can be approximated as follows. 
\begin{equation}
{\cal V}_{hor}(Y)={1\over 3}(Y^IH_I)_{hor}={1\over 3}(Y^I)_{hor}
({q_I\over r^2}).\label{horizon}
\end{equation}
However, $Y_{hor}^I={\cal V}_{hor}^{1/3}(Y)X_{hor}^I$, and thus
\begin{equation}
{\cal V}_{hor}^{2/3}(Y)={1\over 3}{Z_{hor}\over r^2}
\end{equation} 
where $Z=q_IX^I$ is the central charge, and $Z_{hor}$ is its value at 
the horizon. 
Also, equation (\ref{jennie})  which 
defines the moduli over space-time, becomes near the horizon
\begin{equation}
(ZX_I)_{hor}=q_I,
\end{equation}
which is the equation obtained from the extremization of the central charge 
\cite{chams}, $Z_{hor}=Z_{ext}$. 

The ADM mass of black holes in five dimensions is given by \cite{perry}
\begin{equation}
g_{tt}=1-{8G_NM_{ADM}\over 3\pi r^2}+\cdots
\end{equation}
where $G_N$ is Newton's constant. This implies for our metric 
\begin{equation}
{\cal V}(Y)=Y^IY_I=1+{2G_NM_{ADM}\over \pi r^2}+\cdots
\end{equation}
If we expand $Y^I$ as
\begin{equation}
Y^I=Y_{\infty}^I+{\beta^I\over r^2}+\cdots
\end{equation} 
and write the harmonic functions as $H_I=h_I+{q_I\over r^2}$, then one obtains 
\begin{equation}
1+{2G_NM_{ADM}\over \pi r^2}+\cdots={1\over 3}\Big(h_IY^I_{\infty}+
{h_I\beta^I+Y_\infty^Iq_I\over r^2}+\cdots\Big)
\end{equation}
However, from the relation $Y_I\partial_rY^I={1\over3}\partial_r{\cal
  V}(Y)$ one obtains
$\beta^Ih_I={{2\over \pi}G_NM_{ADM}}$, and thus the ADM mass is
related to the central charge by
\begin{equation}
M_{ADM}={\pi\over 4G_N} Z_\infty,
\end{equation}
where we have used $Y^I_{\infty}=X^I_{\infty}$.
Therefore these black holes saturate the BPS bound as should be expected.

The Bekenstein-Hawking entropy ${\bf S}_{BH},$ related to the area of the 
horizon ($r=0$) $\bf A,$ is given by
\begin{equation}
{\bf S}_{BH}={{\bf A}\over 4G_N}=
{\pi^2\over 2G_N}\sqrt{\Big({Z_{hor}\over 3}\Big)^3-
\alpha^2}.
\end{equation}
If one assumes that the values of 
the moduli at the horizon are
valid throughout the entire space-time, then one obtains the double-extreme
black hole solution \cite{chams} 
where the metric takes form
\begin{equation}
ds^2=-\Big(1+{Z\over 3r^2}\Big)^{-2}\Big(dt-
{\alpha\sin^2\theta\over r^2} d\phi+
{\alpha\cos^2\theta\over r^2}\Big)^2+
\Big(1+{Z\over 3r^2}\Big)(dr^2+r^2d\Omega^2).
\end{equation}

As a specific example, consider the so-called $STU=1$ model 
\cite{chaka, sabra2}, $(X^1=S, X^2=T, X^3=U).$
The equations one obtains from (\ref{jennie}) are given by \cite{sabra2}
\begin{eqnarray}
e^{2U}TU&=&H_0,\nonumber\\
e^{2U}SU&=&H_1,\nonumber\\
e^{2U}ST&=&H_2,
\label{judy}
\end{eqnarray}
where $H_0$, $H_1$ and $H_2$ are harmonic functions. Equation (\ref{judy})
together with the fact that $STU=1$ implies the following solution for
the metric and the moduli fields,
\begin{equation}
e^{2U}=(H_0H_1H_2)^{{1\over3}}  
\end{equation}
and
\begin{eqnarray}
S&=&\Big({H_1H_2\over H_0^2}\Big)^{1\over3},\nonumber\\
T&=&\Big({H_0H_2\over H_1^2}\Big)^{1\over3},\nonumber\\
U&=&\Big({H_0H_1\over H_2^2}\Big)^{1\over3},
\end{eqnarray}
If one writes the harmonic function as
$H_I=1+{q_I\over r^2},$
for this model, the ADM mass and the entropy are
\begin{eqnarray}
M_{ADM}&=&{\pi\over 4G_N}({q_0}+{q_1}+{q_2}),\nonumber\\
{\bf S}_{BH}&=&{\pi^2\over 2G_N}\sqrt{{q_0q_1q_2}-
\alpha^2}.
\end{eqnarray}
The  solution obtained is the one found in 
\cite {Tsey, mirjam} in a different context.

In conclusion, we have given an algorithm for obtaining general BPS black 
holes  which breaks half of supersymmety ($\Gamma^0\epsilon=-i\epsilon$)
for the theory of $N=2$ $d=5$ supergravity coupled to an arbitrary number
of vector multiplets. These solutions where expressed in terms of the
rescaled cubic 
polynomial which in the case of Calabi-Yau compactification
corresponds to the intersection 
form. For the solutions found, the gauge 
fields are related to the special coordinates $X^I$ via the relations
(\ref{grace}) and (\ref{mona}). It should be emphasized  that the
unbroken supersymmetry
of the BPS solutions does not fix the configuration completely but rather
 provide a relationship between the various physical fields (metric,
 gauge and scalar fields) 
as well as a constraint on $w$, the function that allows for 
rotating solution. The self-duality of the field 
strength of $w$ forces the angular momentum in the two orthogonal
planes to be equal.
Such a condition was also derived in the conformal sigma model
approach, as arising from the requirement of
conformal invariance \cite{Tsey}. The black hole solution can be fixed
in terms of 
space-time functions by solving for the equations of motion for the 
gauge fields. It is of interest to generalise these solutions to the 
non-extreme case and also
to obtain microscopically  their entropies. 
\vskip 2cm 
{\large \bf Acknowledgment}:  W. A. Sabra was supported by DFG and 
partially by DESY-Zuethen during most of this work, and would like to thank
the Institute of Theoretical Physics at ETH, for hospitality.
\vfill\eject

\end{document}